\documentclass[conference]{IEEEtran}
\usepackage{cite}
\ifCLASSINFOpdf
   \usepackage[pdftex]{graphicx}
\else
\fi
\usepackage{amsmath}
\usepackage{amssymb}
\usepackage{xcolor}
\usepackage[nolist,nohyperlinks]{acronym}

\usepackage{algorithm}
\usepackage[noend]{algpseudocode}
\usepackage{url}
\DeclareMathOperator*{\argmax}{argmax} % no space, limits underneath in displays

\acrodef{UE}{user equipment}
\acrodef{BS}{base station}
\acrodef{Tx}{transmitter}
\acrodef{Rx}{receiver}
\acrodef{CFO}{carrier frequency offset}
\acrodef{ULA}{uniform linear array}
\acrodef{OFDM}{orthogonal frequency division multiplexing}
\acrodef{AWGN}{additive white Gaussian noise}
\acrodef{LO}{local oscillator}
\acrodef{MAP}{maximum \textit{a posteriori}}
\acrodef{mmWave}{millimeter wave}
\acrodef{subTHz}{sub-terahertz}
\acrodef{LOS}{line-of-sight}
\acrodef{NLOS}{non-line-of-sight}
\acrodef{MCMC}{Markov chain Monte Carlo}
\acrodef{VB}{variational Bayes}
\acrodef{EP}{expectation propagation}
\acrodef{KL}{Kullback-Leibler}
\acrodef{SBL}{sparse Bayesian learning}
\acrodef{RV}{random variable}
\acrodef{PDF}{probability density function}
\acrodef{EM}{expectation maximization}
\acrodef{5G}{fifth generation}
\acrodef{iid}{independent identically distributed}
\acrodef{ML}{maximum likelihood}
\acrodef{SCA}{successive convex approximation}
\acrodef{MM}{majorization-minimization}
\acrodef{ADMM}{alternating direction method of multipliers}
\acrodef{SNR}{signal-to-noise ratio}
\acrodef{AoA}{angle of arrival}
\acrodef{AoD}{angle of departure}
\acrodef{MSE}{mean squared error}
\acrodef{RMSE}{root mean squared error}
\acrodef{ReMSE}{relative mean squared error}
\acrodef{PSO}{particle swarm optimization}
\acrodef{NM}{Nelder-Mead}
\acrodef{B5G}{beyond 5G}
\acrodef{6G}{sixth generation standard}
\acrodef{MIMO}{multiple-input multiple- output}
\acrodef{LLN}{law of large numbers}
\acrodef{LGR}{large grid regime}
\acrodef{LPA}{long pilot approximation}
\acrodef{MC}{Monte Carlo}
\acrodef{ToF}{time of flight}
\acrodefplural{TOF}[TOFs]{times of flight}
\acrodef{CACC}{cross-antenna cross-correlation}
\acrodef{MMSE}{minimum mean square error}
\acrodef{DTFT}{discrete-time Fourier transform}
\acrodef{DFT}{discrete Fourier transform}
\acrodef{FFT}{Fast Fourier transform}
\acrodef{2P2T}{two precoders two transmissions}
\acrodef{TVP}{time-varying precoder}
\acrodef{RF}{radio frequency}
\acrodef{PSF}{point spread function}
\acrodef{DBSCAN}{density-based spatial clustering of applications with noise}
\acrodef{SSE}{Sum Square Error}
\acrodef{TDMA}{time division multiple access}
\acrodef{FDMA}{frequency division multiple access}
\acrodef{CDMA}{code division multiple access}
\acrodef{BER}{bit error rate}
\acrodef{MUSIC}{multiple signal classification}
\acrodef{URA}{uniform rectangular array}
\acrodef{JCS}{joint communication and sensing}
\acrodef{NPNT}{N-precoders N-transmissions}
\acrodef{SOR}{successive over-relaxation}
\acrodef{COMPAS}{concurrent mapping, positioning, and synchronization}
\acrodef{SLAM}{simultaneous localization and mapping}
\acrodef{AECD}{alternating exact coordinate descent}
\acrodef{ISAC}{integrated sensing and communication}
\acrodef{JSC}{joint sensing and communication}
\acrodef{FR2}{frequency range 2}
\acrodef{CSI}{channel state information}
\acrodef{SAGE}{space-alternating generalized expectation-maximization}
\acrodef{PCE}{parametric channel estimation}
\acrodef{WSNSAP}{wideband spatial nonstationary wireless channels with antenna polarization}
\acrodef{GN}{Gauss-Newton}
\acrodef{RiMAX}{Richter's maximum likelihood estimation}
\acrodef{CP}{CANDECOMP/PARAFAC}
\acrodef{MSVD}{multilinear singular value decomposition}
\acrodef{ESPRIT}{estimation of signal parameters via rotational invariant techniques}
\acrodef{BIC}{Bayesian information criteria}
\acrodef{AIC}{Akaike information criterion}
\acrodef{WAN}{wide area network}
\acrodef{MAN}{metropolitan area network}
\acrodef{LAN}{local area network}
\acrodef{MHR}{multidimensional harmonic retrieval}
\acrodef{SIC}{successive interference cancellation}
\acrodef{NOMA}{non-orthogonal multiple access}

\usepackage{subcaption}
\usepackage{diagbox}

\begin{document}

%\addtolength{\textfloatsep}{-0.1in}

\title{Doppler-Robust Maximum Likelihood Parametric Channel Estimation for Multiuser MIMO--OFDM}

\author{\IEEEauthorblockN{Enrique~T.~R.~Pinto and Markku~Juntti}
\IEEEauthorblockA{Centre for Wireless Communications (CWC), University of Oulu, Finland\\
\{enrique.pinto, markku.juntti\}@oulu.fi}}

\markboth{Journal of \LaTeX\ Class Files,~Vol.~14, No.~8, August~2015}%
{Shell \MakeLowercase{\textit{et al.}}: Bare Demo of IEEEtran.cls for IEEE Journals}




\maketitle

\noindent
\begin{abstract}
    The high directionality and intense Doppler effects of millimeter wave (mmWave) and sub-terahertz (subTHz) channels demand accurate localization of the users and a new paradigm of channel estimation. For orthogonal frequency division multiplexing (OFDM) waveforms, estimating the geometric parameters of the radio channel can make these systems more Doppler-resistant and also enhance sensing and positioning performance. In this paper, we derive a multiuser, multiple-input multiple-output (MIMO), maximum likelihood, parametric channel estimation algorithm for uplink sensing, which is capable of accurately estimating the parameters of each multipath that composes each user's channel under severe Doppler shift conditions. The presented method is one of the only Doppler-robust currently available algorithms that does not rely on line search.
\end{abstract}

\noindent

% \fi
\IEEEpeerreviewmaketitle


\section{Introduction}
\IEEEPARstart{H}{igher} frequency ranges such as \ac{mmWave} and \ac{subTHz} have been the target of intensive research recently due to their attractive properties for many mobile radio use cases. The ample availability of spectrum in these spectra is considered to be a major enabler for the desired \textit{Tbps} rates \cite{6g_spectrum}. Beyond throughput, larger bandwidths also allow improved sensing and positioning performance by decreasing the \ac{ToF} uncertainty. The radio channels at \ac{mmWave} and \ac{subTHz} are also convenient for localization and sensing, since they are quasi-optical, meaning that most of the power is transferred through \ac{LOS} and low-order reflections, and diffraction and high-order reflections are not as significant \cite{mmWave_prop1}. Accurate localization is fundamental at these frequency ranges due to the high channel directionality, which is a consequence of massive \ac{MIMO} arrays, and to the significant effects of Doppler shifts, which are proportional to the carrier frequency. This means that medium to high mobility channels have a short coherence time (smaller than $100~\mu s$) and that the usual channel estimation procedures are not sufficiently effective, since channel estimates quickly become outdated. The ability to perform sensing and localization using the mobile communications infrastructure, i.e., \ac{JSC}, is also an attractive perspective that will simultaneously augment the communications performance and may provide vital information for other applications.\par

Once that directly estimating the channel matrix/tensor is not sufficient for high-mobility \ac{mmWave} and \ac{subTHz} channels, performing \ac{PCE} becomes necessary. By \ac{PCE} it is meant that the estimation procedure can extract the multipath components that make up the radio channel as well as their parameters such as amplitude, phase, \ac{ToF}, \ac{AoA}, \ac{AoD}, and Doppler shift. One of the earliest methods for this application is the celebrated \ac{SAGE} procedure \cite{sage}, which maximizes the likelihood function of the received signal. While being the state of the art tool in offline channel modelling and propagation characterization, \ac{SAGE} is known to not fit well for real-time applications, specially due to its coordinate-wise updating with exhaustive line-search. More recently, the \ac{SAGE} algorithm has been extended by Zhou \textit{et al.} \cite{sage_nearfield} with the SAGE \ac{WSNSAP} algorithm. Also, another popular maximum likelihood method is the \ac{RiMAX}, which is a specialization of the \ac{GN} algorithm. \par

The tensor decomposition methods from an alternative approach to the maximum likelihood estimation. In \cite{parafac_zhou, henk_cp}, decompositions such as the \ac{CP} decomposition and the \ac{MSVD} \cite{tensorreview} are used to estimate the channel parameters. While these methods are generally accurate and fast, they require first estimating the channel tensor, on which the tensor decompositions will then be performed. This is a problem because pilot-based \ac{MIMO} channel estimation requires the channel to remain approximately constant for at least $N_t$ symbols, where $N_t$ is the number of transmit antennas, which requires a very fast symbol period due to the short coherence time.\par

In this paper, we introduce a  maximum likelihood method for multiuser, parametric \ac{OFDM} channel estimation for uplink sensing. The proposed procedure can  estimate reliably the channel parameters using measurements that span several coherence time intervals, yielding accurate estimates for the multipath magnitudes, phases, \acp{ToF}, \acp{AoA}, \acp{AoD}, and Doppler shifts. The procedure also iteratively estimates the number of multipaths using information theoretic criteria, such as a generalization of the \ac{AIC} \cite{akaike}.%, and uses it as a stopping criterion.
In Section \ref{sec:model}, we introduce the model considered in this paper. Then, in Section \ref{sec:estimation}, we present the estimation framework and introduce the background for the algorithm shown in Section \ref{sec:opt}. Finally, we analyse some numerical results in Section \ref{sec:numres} and make our concluding remarks in Section \ref{sec:conclusion}.

\section{System Model}\label{sec:model}
Consider the following uplink multiuser \ac{OFDM} received signal model \cite{zhang_enabling_jsc}
\vspace{-0.3cm}
\begin{multline} 
    \mathbf{y}_{nt} = \sum^K_{k=1} \sum^{L_k}_{\ell=1} b_{\ell k} e^{-j2\pi n (\tau_{\ell k}+\tau_{o k})f_\text{scs}} e^{j2\pi t (f_{\ell k} + f_{o k}) T_s} \\
    \cdot \mathbf{a}(\phi_{\ell k}) \mathbf{a}^T(\theta_{\ell k}) \mathbf{x}^{k}_{nt} + \mathbf{w}_{nt}, \label{eq:rec_sig}
\end{multline}
where $n$ and $t$ denote the \ac{OFDM} subcarrier and symbol index, respectively; $\mathbf{y}_{nt}$ is the signal received by the \ac{BS} at the $n$th subcarrier and $t$th symbol; $L$ is the number of multipath components; $K$ is the number of active users; symbol $k$ is user index; the $\ell$ index indicates the path; $b$ is the path gain; $\tau$ is the propagation delay; $\tau_{o}$ is the clock timing offset between the \ac{UE} and the \ac{BS}; $f_{\text{scs}}$ is the subcarrier spacing $B/N_c$, where $B$ is the bandwidth; $f$ is the Doppler frequency; $f_{o}$ is the \ac{CFO} of between \ac{UE} and the \ac{BS}; $T_s$ is the \ac{OFDM} symbol length; $\mathbf{a}(\phi/\theta)$ is the \ac{ULA} response vector with $N_r$/$N_t$ antennas and angle of arrival/departure $\phi/\theta$, given by $\mathbf{a}(\phi/\theta) = \begin{bmatrix} 1 & e^{-j \pi \sin(\phi/\theta)} & \cdots & e^{-j\pi (N_{R/T} -1) \sin(\phi/\theta)} \end{bmatrix}^T$, where ``$\phi/\theta$" here denotes ``either $\phi$ or $\theta$"; $\mathbf{x}^k_{nt}$ is the transmitted pilot of the user $k$ at the $n$th subcarrier and $t$th symbol; and finally $\mathbf{w}_{nt}$ is \ac{AWGN} at the $n$th subcarrier and $t$th symbol with covariance $N_0 \mathbf{I}_{N_r}$. Because the signal is transmitted by the \ac{UE}, this scenario is called uplink sensing. The model in (\ref{eq:rec_sig}) assumes symbol-level synchronization between the users and the \ac{BS}, such that the \ac{OFDM} resource grids approximately align and the transmitted symbols of each user at each $(n,t)$ pair are known. \par

\begin{figure}[!htbp]
    \centering
    \includegraphics[width=\linewidth]{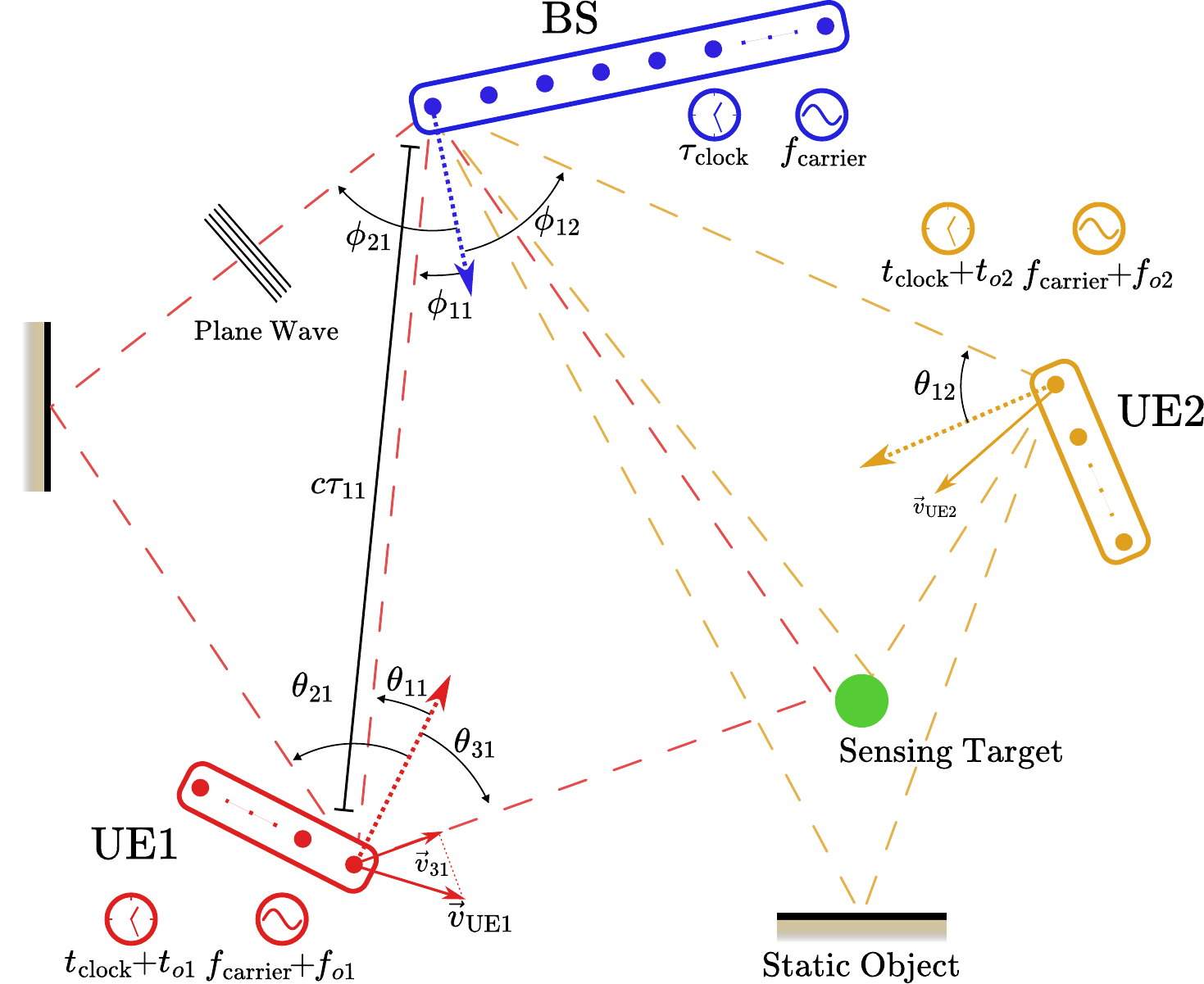}
    \caption{Representation of the considered channel model. The \ac{UE}1 moves with velocity $\vec{v}_\text{UE1}$, which produces Doppler shifts proportional to the projection of the velocity vector on the direction of departure of the paths; an example is provided for path $(\ell,k)=(3,1)$. The clock and local oscillator offsets are also represented. The \ac{BS} is assumed to be in the far-field. The environment is represented by the ``Static Object" elements and a possible sensing target is depicted as a green circle. For clarity, only some multipath elements are labeled.}
    \label{fig:geometry}
\end{figure}

% We rewrite the $u$th entry of the array response vector $\mathbf{a}_n(\phi)$ in the following fashion
% \begin{equation}
%     [\mathbf{a}_n(\phi)]_u = \exp\left(-j2\pi u f_n \sin(\phi)\right),\; f_n = \frac{(f_{\text{min}} + n f_\text{scs} ) d}{c}\label{eq:array_response_subcarrier}
% \end{equation}
% where $f_\text{min}$ is the frequency of the ``lowest" subcarrier, $c$ is the speed of light. The above expression details explicitly the relation of the array response with the subcarrier index $n$. A similar expression holds for $[\mathbf{a}_n(\theta)]_u$.\par

Far-field models are typically sufficient for \ac{MAN} or \ac{WAN} contexts in the uplink direction due to the reduced dimensions of the transmit antenna. Even at \ac{mmWave} and \ac{subTHz} bands, the Fraunhofer distance, which defines the soft boundary between the near and far fields, is only around a couple of meters. Near-field models are nonetheless important and are probably necessary for uplink \ac{LAN} deployments (and possibly for \ac{MAN} as well).\par

Estimating the offsets from (\ref{eq:rec_sig}) is not possible without additional assumptions. Therefore, we group the offsets with the path parameters to avoid estimation ambiguity by defining $\omega_{1\ell k} =-2\pi(\tau_{\ell k}+\tau_{o k})f_\text{scs}$ and $\omega_{2 \ell k} = 2\pi(f_{\ell k} + f_{o k}) T_s$. In this work, we do not tackle the estimation of the offsets, instead we focus exclusively on estimating $\boldsymbol{\xi}_{\ell k} = (b_{\ell k}, \omega_{1 \ell k}, \omega_{2 \ell k}, \phi_{\ell k}, \theta_{\ell k}) \forall \ell, k$. Additional estimation methods would be required to identify the offsets.
% One integrated approach for the estimation of the time offset as well as the reflector and \ac{UE} positions has been proposed by Mendrzik \textit{et al.} \cite{Mendrzik} by means of a message passing-based inference engine called \ac{COMPAS}. The slightly different inference problem of multipath-based \ac{SLAM} is addressed in \cite{henk_vehicular_slam, diffuse_mp_slam}.\par

% We intend to perform a \ac{MAP} sequential estimation procedure to extract $\boldsymbol{\xi}_\ell$ for as many paths as possible, while avoiding misdetections. The inference procedure will be detailed in the following section.\par
\vspace{-0.1cm}
\section{Parameter Update Framework} \label{sec:estimation}
\vspace{-0.1cm}

Define $\mathbf{y}=\text{vect}(y_{ntu})$, where $\text{vect}(\cdot)$ denotes the tensor vectorization operation, also denote by $\boldsymbol{\xi}$ the vector of sensing parameters $\boldsymbol{\xi}_{\ell k}$ for all detected paths and all users, then the maximum likelihood estimate of $\boldsymbol{\xi}$ given the data $\mathbf{y}$ is given by
\begin{equation}
    \hat{\boldsymbol{\xi}} = \argmax_{\boldsymbol{\xi}} p(\mathbf{y}|\boldsymbol{\xi}) = \argmax_{\boldsymbol{\xi}} \prod_{ntu} p(y_{ntu}|\boldsymbol{\xi}),
\end{equation}
The conditional \ac{PDF} of the data is complex normal $ y_{ntu}|\boldsymbol{\xi} \sim \mathcal{CN}\left(\mu_{ntu}(\boldsymbol{\xi}) ,N_0 \right)$, where the mean is given by 
\vspace{-0.1cm}
\begin{equation}
    \mu_{ntu} = \sum^K_{k=1} \sum^{L_k}_{\ell=1} b_{\ell k} e^{j \omega_{1\ell k} n } e^{j \omega_{2\ell k} t} e^{-j \pi u \sin(\phi_\ell)} \mathbf{a}^T(\theta_{\ell})\mathbf{x}_{nt}, \label{eq:mean}
\end{equation}
where the dependence on $\boldsymbol{\xi}$ has been omitted. We then write the estimation as a constrained minimization problem
\vspace{-0.1cm}
\begin{align}
       \min_{\boldsymbol{\xi}} & \frac{1}{N_0}\sum_{ntu} \left| y_{ntu} - \mu_{ntu}(\boldsymbol{\xi}) \right|^2  \label{eq:opt_obj}\\
       \text{s.t. } &\angle b_{\ell k},\, \omega_{1\ell k},\, \omega_{2\ell k} \in (-\pi.\pi);\; \phi_{\ell k},\, \theta_{\ell k} \in \left( -\frac{\pi}{2}, \frac{\pi}{2} \right) \; \forall \ell, k. \notag
\end{align}
The objective function is nonconvex over $\boldsymbol{\xi}$ and is quite high-dimensional. Local descent methods, such as gradient descent and its variations, are thus not very effective. Furthermore, the computational cost for objective function evaluation makes many global optimization methods, such as particle swarm and simulated annealing, not viable  for real-time applications. One technique that is successful for this problem is an augmented form of \ac{AECD}.\par

Because (\ref{eq:opt_obj}) is convex in $b_{\ell k}$, a closed form solution exists, given by
\begin{equation}
    b_{\ell' k'}(\boldsymbol{\xi}_{\ell' k'}) = \frac{\sum_{n,t,u} \alpha^{uk'*}_{\ell'nt} \left( y^u_{nt} - \sum_{(\ell,k) \neq (\ell',k')} b_{\ell k} \alpha^{uk}_{\ell nt} \right)}{\sum_{n,t,u} |\alpha^{uk'}_{\ell'nt}|^2}.
    \label{eq:b_opt}
\end{equation}
We propose estimating one path at a time in alternating fashion by substituting (\ref{eq:b_opt}) into the corresponding path in (\ref{eq:opt_obj}), while keeping all the other parameters $\boldsymbol{\xi}_{\ell k}$, for $(\ell,k)\neq(\ell',k')$, fixed. By substituting $b_{\ell}$, estimating the path coefficient becomes a consequence of accurately estimating the other parameters. We flexibly denote by $f(\boldsymbol{\xi}_{\ell' k'})$ the objective function with $b_{\ell' k'}$ substituted and the other paths and users kept fixed.

The exact coordinate descent requires that the gradient along that coordinate direction is zero. We will show that the partial derivatives of the log-likelihood term with relation to $\theta_{\ell k}$, $\phi_{\ell k}$, $\omega_{1\ell k}$, and $\omega_{2\ell k}$, are given by the Fourier series over each respective parameter. The roots of the resulting series are candidate solutions for the coordinate update. We can solve for the roots of the Fourier series by converting it into a companion matrix eigenvalue problem and applying a transformation to the computed eigenvalues \cite{fourier_roots}. Finally, we evaluate the objective on the roots and select the smallest one.\par
We now present the partial derivatives of $f(\boldsymbol{\xi}_{\ell' k'})$ over the $\omega_{1\ell' k'}$, $\omega_{2\ell' k'}$, $\theta_{\ell' k'}$, and $\phi_{\ell' k'}$ coordinates. We omit the derivation due to space constraints. Over the following section, some indices will be moved from the subscript to superscript in order to save space. Additionally we denote the transmitted signal of user $k$ at transmit antenna $v$ as $x^{kv}_{nt}$.

\vspace{-0.2cm}
\subsection{Partial Derivative Over $\omega_{1\ell'k'}$, $\omega_{2\ell'k'}$, and $\phi_{\ell' k'}$}
The partial derivative over $\omega_{1\ell'k'}$ is a Fourier series indexed over $m\in[1-N_c,\dots,0,\dots,N_c-1]$ with coefficients
\begin{multline}
    \hat{c}_m = jm\Bigg[(c_m \ast c^*_{-m})\left(\sum_{n,t,u} |\mathbf{a}^T(\theta_{\ell'k'})\mathbf{x}^{k'}_{nt}|^2 \right) \\
    + \text{vec}\left( \sum_{n,t,u}  c_{m-n}d^u_{nt} + c^*_{n-m}d^{u*}_{n,t} \right) \Bigg]_m,
\end{multline}
where ``$\ast$" denotes discrete convolution, $\text{vec}_m(\cdot)$ means putting the elements of the argument in a coefficient vector properly indexed over $m$, and $[\cdot]_m$ means taking the $m$th element of the vector. Furthermore
\begin{gather*}
    y^{uk}_{\ell,n,t} = b_{\ell k} e^{j \omega_{1\ell k} n } e^{j \omega_{2\ell k} t} e^{-j\pi u \sin(\phi_{\ell k})} \mathbf{a}^T(\theta_{\ell k})\mathbf{x}^k_{n,t} \\
    a^{uk'}_{\ell'nt} = \!\!\!\!\!\!\!\!\! \sum_{(\ell,k)\neq(\ell',k')} \!\!\!\!\!\!\! y^{uk}_{\ell,n,t}\\
    \Bar{\alpha}^{uk,*}_{n,t,u} = e^{j\omega_{2\ell k}t} e^{-j\pi u \sin(\phi_{\ell k})} \mathbf{a}_n^T(\theta_{\ell k}) \mathbf{x}^k_{n,t} \\
    c_{-m} = \!\!\frac{\sum_{t,u}\Bar{\alpha}^{uk',*}_{m,t,u}\left( y^u_{m,t} - a^{uk'}_{\ell',m,t} \right)}{\sum_{n,t,u}|\mathbf{a}_n^T(\theta_\ell')\mathbf{x}_{n,t}|^2}; \\
     d^u_{n,t} = \Bar{\alpha}^{uk',*}_{n,t,u} \left(y^u_{n,t} - a^{uk'}_{\ell',n,t} \right)^*
\end{gather*}
The partial derivative over $\omega_{2\ell'k'}$ is similar, by symmetry.
We can also see that the partial derivative over $-\pi\sin(\phi_{\ell'})$ follows similarly. The derivative over $\sin(\phi_{\ell'})$ is obtained by
\begin{equation}
    \frac{\partial f(\boldsymbol{\xi})}{\partial \sin(\phi_{\ell'})}  = -\pi \frac{\partial f(\boldsymbol{\xi})}{\partial -\pi\sin(\phi_{\ell'})},
\end{equation}
while also doing the appropriate variable exchanges to preserve the symmetry.

\subsection{Partial Derivative over $\sin(\theta_{\ell'k'})$}
For $\theta_{\ell'k'}$, we take the derivative over $\sin(\theta_{\ell'k'})$ and exploit the bijectivity of the sine function over the $(-\frac{\pi}{2},\frac{\pi}{2})$ range to compute the value of $\theta_{\ell'k'}$ that satisfies $\frac{\partial f(\boldsymbol{\xi}_{\ell'k'})}{\partial\sin(\phi_{\ell'})}=0$ with smallest objective value. For the resulting derivative to be a Fourier series, the transmitted signal must satisfy
\vspace{-0.1cm}
\begin{equation}
    \frac{\partial}{\partial \theta_{\ell'k'}} \sum_{n,t}|\mathbf{a}^T(\theta_{\ell'k'})\mathbf{x}_{n,t}|^2 = 0. \label{eq:isotropy}
\end{equation}
otherwise the product rule with $b(\boldsymbol{\xi}_{\ell'k'})$ breaks the Fourier series structure. We refer to a signal satisfying (\ref{eq:isotropy}) as \textit{isotropic}, because the total transmitted power is independent of the angle.
The derivative of $f$ with respect to $\sin(\theta_{\ell'})$ has coefficients indexed over $m \in [1-2N_t,\dots,0,\dots,2N_t-1]$ given by
\vspace{-0.1cm}
\begin{equation}
    \hat{q}_m = j \pi m \, \Bigg[ \sum_{n,t,u} q^m_{nt} \ast q^{-m,*}_{nt} + \text{vec} \left( q^m_{nt} \hat{a}^{uk'}_{\ell'nt} + q^{-m,*}_{nt} \hat{a}^{uk'}_{\ell'nt} \right) \! \Bigg]_m \!\!\!\!\!,
\end{equation}
in which
\begin{gather}
    q^0_{n,t} \! = \!\! \sum^{N_t-1}_{v=0} \Bar{x}^v_{\ell'k'} x^{k'v}_{nt}; \;
    q^m_{n,t} \! = \!\! \begin{cases}
         \sum^{N_t-1}_{v=m} \Bar{x}^v_{\ell'k'} x^{k',v-m}_{nt}, m>0 \\
         \sum^{N_t-1}_{v=-m} \Bar{x}^{v+m}_{\ell'k'} x^{k'v}_{nt}, m<0
    \end{cases}\\
    \Bar{\mathbf{x}}_{\ell'k'} = \sum_{n,t,u} \mathbf{x}^{k',*}_{nt}\Bar{\alpha}^{uk',*}_{\ell'nt} \frac{y^u_{n,t} - a^{uk'}_{\ell'nt}}{N_R\sum_{n,t} |\mathbf{a}^T(\theta_{\ell'k'})\mathbf{x}^k_{nt}|} \\
    \Bar{\alpha}^{uk'}_{\ell'nt} =  e^{j \omega_{1\ell k'} n } e^{j \omega_{2\ell k'} t} e^{-j\pi u \sin(\phi_{\ell k'})} \\
    \hat{a}^{uk'}_{\ell'nt} = \Bar{\alpha}^{uk'}_{\ell'nt} \left(y^u_{nt} - a^{uk'}_{\ell'nt} \right)^*,
\end{gather}
where $\Bar{x}^v_{\ell'k'}$ denotes the $v$th element of $\Bar{\mathbf{x}}_{\ell'k'}$, and $\Bar{\alpha}^{uk'}_{\ell'nt}$ has been redefined for convenience.

\section{Optimization Procedure} \label{sec:opt}
A high level description of the proposed estimation algorithm is presented in Algorithm \ref{alg:main}. We omit some details for space constraints, but provide a short description of the steps. \par

For the optimization problem at hand, the gradient or coordinate descent methods by themselves are ineffective in providing acceptable solutions. Thus, we augment the coordinate descent procedure with a combination of momentum and a \ac{SOR} update, which is effective in escaping local optima and improving the estimation results. The $m$th update of an arbitrary parameter $\xi$ is given by
\begin{equation}
    \xi_{m+1} = \text{Wrap}_{\xi}\left( (1-\rho)\xi_{m} + \rho \Hat{\xi}_{m+1} \right), \label{eq:overrelax_update}
\end{equation}
where $\text{Wrap}_{\xi}(\cdot)$ denotes wrapping the argument value to the valid domain of the parameter, e.g., $\phi$ and $\theta$ should be wrapped to the interval $(-\frac{\pi}{2},\frac{\pi}{2})$ and $\omega_1$ and $\omega_2$ to $(-\pi,\pi)$. We denote the candidate update of $\xi$ at iteration $m$ by $\Hat{\xi}_{m+1}$, this is some function of the output of the exact coordinate descent step. Typically, over-relaxation or under-relaxation are not effective by themselves, and may even be worse than when $\rho=1$. Thus, we propose augmenting the over-relaxed exact coordinate descent with momentum, yielding the following candidate update for each coordinate
\begin{equation}
    \Hat{\xi}_{m+1}  = \xi^{\text{opt}}_m + \eta_m (\xi_m - \xi_{m-1}), \label{eq:momentum_update}
\end{equation}
which is then substituted in (\ref{eq:overrelax_update}) to produce the $m$th update of $\xi$. A path update consists of updating its coordinates one at a time with (\ref{eq:overrelax_update}), and then computing $b_{\ell'k'}$ using (\ref{eq:b_opt}). \par

The channels can be estimated by progressively adding paths. Paths are are updated until convergence, after which another path can be added to the pool of active paths. The addition of a path to user $k$ is considered to have a significant enough contribution to the improvement of the objective function if decreases the generalized \ac{AIC} \vspace{-0.1cm}
\begin{equation}
    AIC_k(L) = \frac{1}{N_0}f(\boldsymbol{\xi}^k_{1:L}) + \gamma_{\text{AIC}} L, \label{eq:aic_eq} \vspace{-0.3cm}
\end{equation}
where $\boldsymbol{\xi}^k_{1:L}$ denotes the parameters of user $k$ up to path $L$ sorted over $\ell$ in descending order of $|b_{\ell k}|$ for each user, and $f(\boldsymbol{\xi}^k_{1:L})$ denotes taking the objective function with respect to only the $k$th user while keeping the others constant. We stop adding paths to a user if adding paths has failed to decrease the \ac{AIC} for a total of $m^\text{max}_{\text{AIC}}$ times. The algorithm stops when the maximum number of outer iterations has been reached, or when the objective has reached a lower threshold which represents optimality. \par

Each user is estimated progressively and in cyclic fashion. This means that we first estimate user 1 until the \ac{AIC} criterion is achieved or $L_\text{max}$ has been reached. Then, the other users are estimated in the same way up to user $K$. The cycle now repeats and user 1 is estimated again. At each new full cycle, the parameters $\boldsymbol{\xi}_k$ of the currently estimated user are cleared to zero, this leads to better results and convergence. Clearing the previous estimates is somewhat unintuitive, but information from those values is still indirectly retained in the estimates of the other users, which considered those (now cleared) parameters for estimation. \par

When estimating user $k$, the paths $(\ell,k)$ are added in an outer loop until convergence. The path update happens in an inner loop, optimization should always start with the newest added path, the remaining paths are updated from the oldest to the newest, this is repeated in cyclic order. For example, if a total of 3 paths is active, the update order follows: $(3,k)$, $(1,k)$, $(2,k)$, cyclically. If a path update has not decreased the objective sufficiently, or if the relative change in the variables was small, then we stop updating this path in the inner loop. The inner loop stops when all the updateable paths have been halted or when a maximum number of inner loop iterations has been reached. We may keep a moving window of the last $L_\text{window}$ paths to avoid having to update all paths every time. When $L_\text{window}$ is properly chosen, this effectively saves computational effort without significant impact on the optimization results.

After the algorithm has stopped, the total number of paths must be estimated. We define the \ac{AIC} tensor with $K$ indices going from 1 to $L_\text{max}$ as
\begin{equation}
    AIC(L_1,\dots,L_K) = \frac{1}{N_0}f(\boldsymbol{\xi}^k_{1:L}) + \gamma_{\text{AIC}} \sum^K_{k=1} L_k. \label{eq:AIC_tensor}
\end{equation}
The estimated number of paths $\mathbf{L}_\text{est}$ is the tuple that minimizes (\ref{eq:AIC_tensor}). \par

\begin{algorithm}[!ht]
\caption{Overview of the main estimation algorithm.}
\label{alg:main}
\begin{algorithmic}[1]
    \Procedure{Main}{$\mathbf{y}$, $\mathbf{x}$, $L_{\text{max}}$}  
        %\State Initialize active paths $L_\text{vec} = [\quad]$;
        \For{$k = [1,\dots, K,1,\dots,K]$}
            \State Initialize $\boldsymbol{\xi}_k=\mathbf{0}$;
            \State Initialize empty path list;
            \For{$L = [1,\dots, L_{\text{max}}]$} 
                \State Add path $(L,k)$ to path list;

                \For{$\text{it}=1,\dots,\text{it}_\text{max}$}
                
                    \For{$(\ell,k)$ in path list (w/ correct order)}
                        \If{Path $(\ell,k)$ is no longer active}
                            \State Skip this path;
                        \EndIf
                        
                        \State Optimize variables $\boldsymbol{\xi}_{\ell k}$ of path $(\ell,k)$;
                        \If{obj. or var. change was small}
                            \State Set $(\ell,k)$ as inactive;
                        \EndIf
                        
                        \If{all paths are inactive}
                            \State Break the ``it" loop; 
                        \EndIf
                    \EndFor
                \EndFor
            \EndFor
    
            \State Evaluate the $\text{AIC}_k \forall k$;
            \If{failed to improve AIC for $m^\text{max}_{\text{AIC}}$ times}
                \State \textbf{continue}; \Comment{Move to next user if current user failed to improve $\text{AIC}_k$ for a total of $m^\text{max}_{\text{AIC}}$ times}
            \EndIf
        \EndFor 
    
        \State Estimate $\mathbf{L}_\text{est}$ with (\ref{eq:AIC_tensor});
        \State \Return $(\boldsymbol{\xi}, \mathbf{L}_\text{est})$; 
    \EndProcedure
\end{algorithmic}
\end{algorithm}
\section{Numerical Results}\label{sec:numres}
In this section, we evaluate the performance of the proposed algorithm with a numerical simulation in which, for simplicity, we consider only the 2 user case. The presented scenario is a Monte Carlo simulation in which the transmit power of user 1 is varied while user 2 is kept at the constant power of $-40$~dBW. The F1 score and the mean absolute error of the parameters each path are presented as a function of the transmit power of user 1. \par

To avoid a detailed and lengthy discussion on the intricacies of \ac{mmWave} and \ac{subTHz} channel modeling, we generate the simulation data as a generic \ac{MHR} problem. By this we mean that the ground truth harmonic frequencies $(\omega_{1\ell k}, \omega_{2\ell k}, \phi_{\ell k}, \theta_{\ell k})$ are just extracted from a uniform distribution with no intention of trying to represent an underlying physical channel. Explicitly, $\omega_1$ and $\omega_2$ use $\mathcal{U}(-\pi,\pi)$ while $\phi$ and $\theta$ use $\mathcal{U}\left(-\frac{\pi}{2},\frac{\pi}{2}\right)$; the path coefficient complex phase $\angle b_{\ell k}$ is also drawn from $\mathcal{U}(-\pi,\pi)$. The path coefficient magnitudes $b_{\ell k}$ are sampled from a distribution with non-negative support, we use a Rice distribution with non-centrality parameter $10^{-2}$ and scale parameter $5\cdot 10^{-3}$ (this obviously does \textit{not} mean that the channel is Rician). The largest path coefficient for each user is multiplied by 1.5 to simulate a \ac{LOS} component. We consider $L_1=L_2=3$, $N_c=30$ subcarriers, $N_s=15$ \ac{OFDM} symbols, $N_r=32$ receive antennas and $N_t=4$ transmit antennas. \par
\begin{figure}[!htbp]
    \centering
    \includegraphics[width=0.9\linewidth]{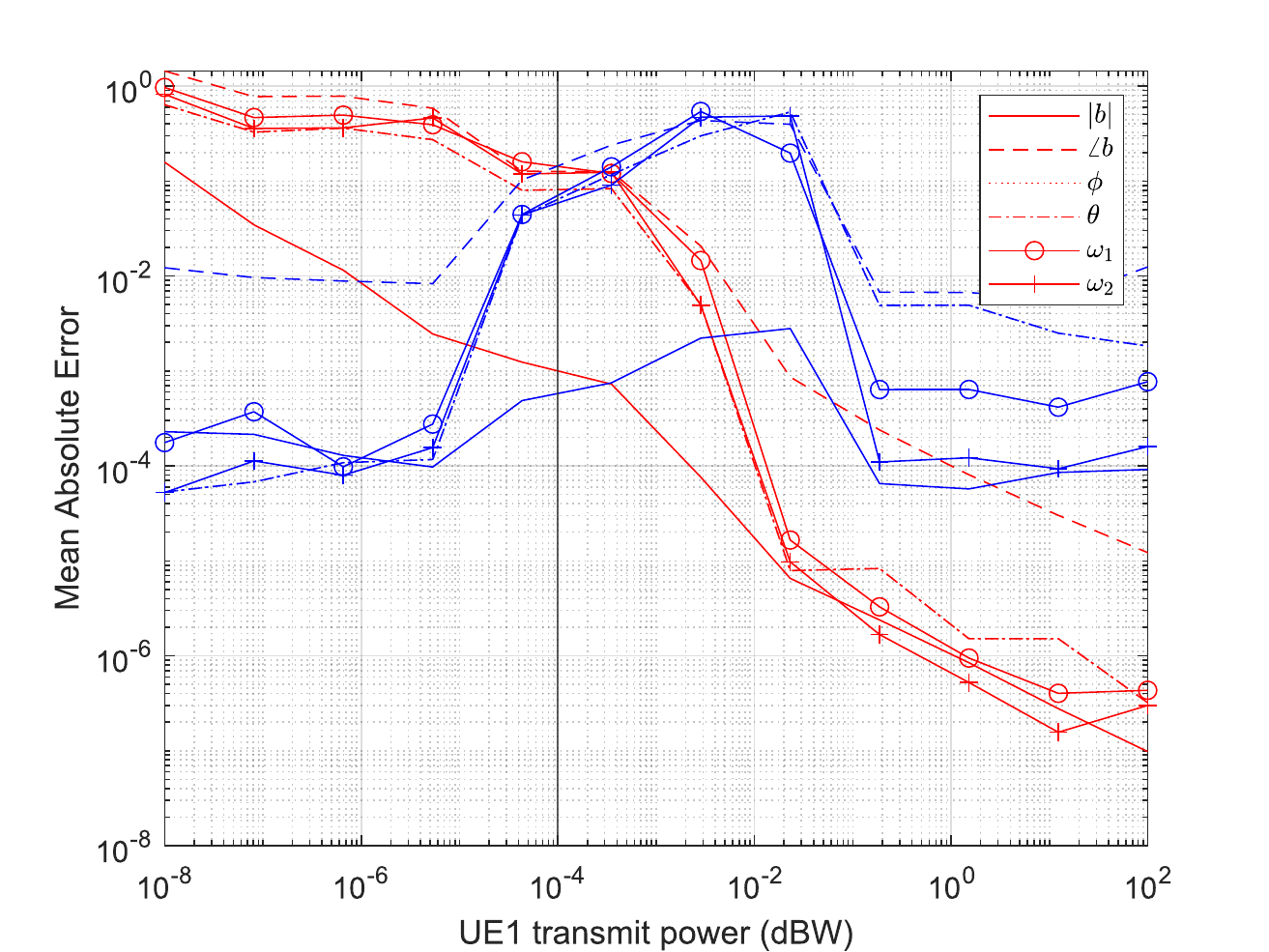}
    \caption{Mean absolute error of path estimates as a function of the transmit power of user 1 (red). The user 2 (blue) transmit power is indicated by the vertical black line.}
    \label{fig:snr_sweep}
\end{figure}

\begin{figure}[!htbp]
    \centering
    \includegraphics[width=0.9\linewidth]{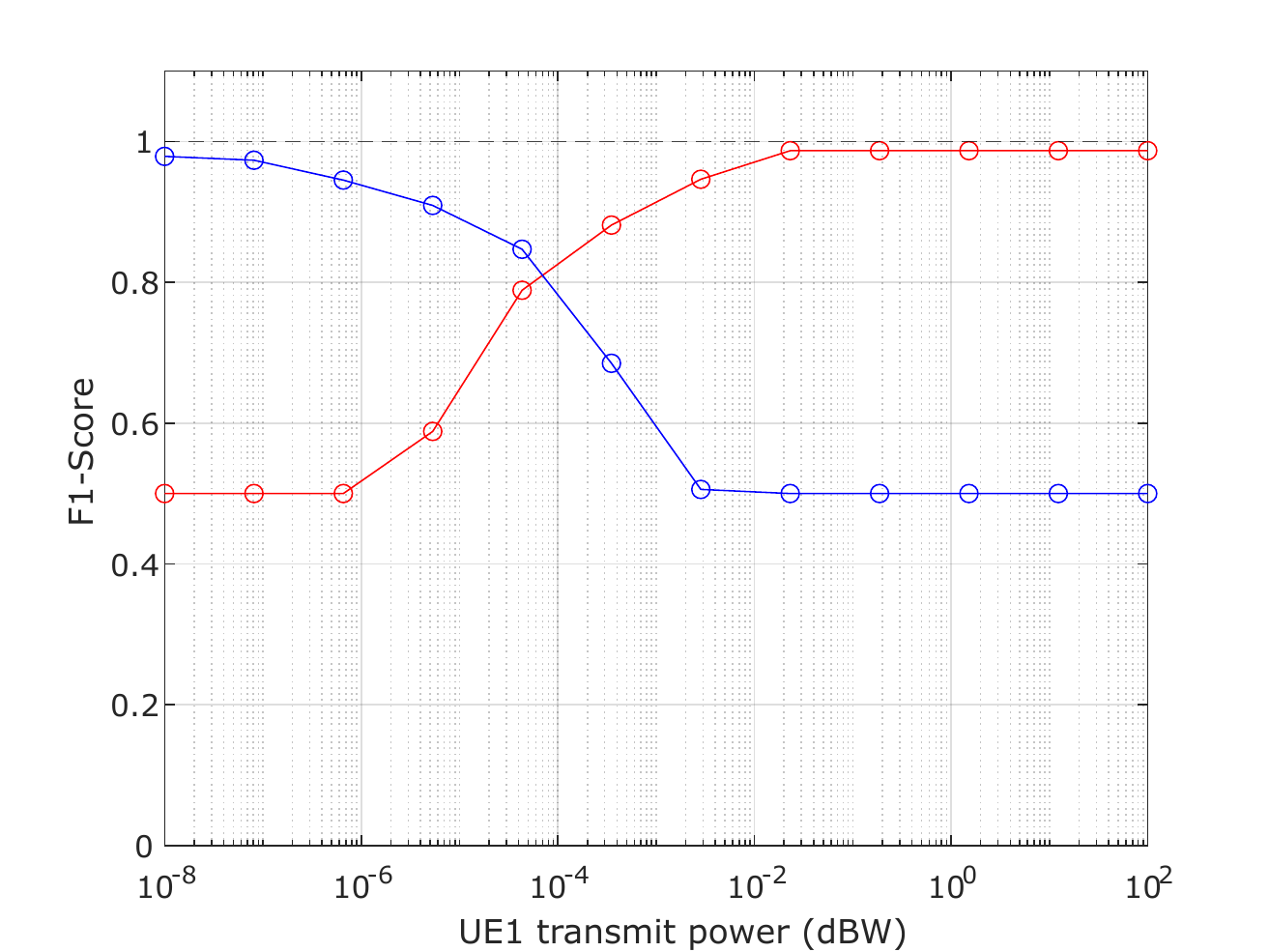}
    \caption{F1 score for the path detection performance of each user as a function of the transmit power of user 1.}
    \label{fig:f1_score}
\end{figure}

Regarding estimator parameters, the initial momentum coefficient is set to $\eta_{\ell k} = 0.1$ and is multiplied by $0.5$ at each time that path is estimated. The momentum and its coefficients are reset whenever the user is estimated again. The over-relaxation parameter is $\rho = 1.05$, the maximum number of inner iterations is $\text{it}_\text{max}=30$, and the maximum \ac{AIC} failures is $m^\text{max}_\text{AIC}=2$. As stopping parameters, the relative change in all path parameters must me smaller than $10^{-8}$ or the objective change must be smaller than $10^{-10}k_\text{Obj}$, where $\gamma_\text{Obj} = \left( \frac{1}{N_0} \sum_{n,t,u} y_{ntu}\right) - N_c N_s N_r$. The users are estimated a total of 3 times, i.e., $k$ iterates through [1 2 1 2 1 2]. \par

The achieved results can be observed in Figures \ref{fig:snr_sweep} and \ref{fig:f1_score}, in which user 1 is represented by red lines and user 2 by blue lines. In both figures, each data point is averaged over 32 iterations. Figure \ref{fig:snr_sweep} presents the absolute error of the estimate of each parameter, averaged across the detected paths. We can see that the estimation performance is greatly deteriorated when both users have similar received powers at the \ac{BS}. This is consistent with the theory of \ac{SIC} in \ac{NOMA}, since it is impossible to decode either user due to the significant interference. When the user 1 transmit power is significantly larger than user 2, it is possible to decode both users with decent performance, because user 1 gets estimated first, which makes way for the estimation of user 2. When the user 2 power is larger than user 1, the estimation error of user 1 is high, which indicates that the quality of the estimation of user 2 is not sufficient to properly cancel its interference. The results from Figure \ref{fig:f1_score} are also intuitive, as the user with higher transmit power experiences the superior path detection performance.

\section{Conclusion}\label{sec:conclusion}
We have introduced a multiuser parametric \ac{OFDM} channel estimation method that is capable of operating with channels of arbitrarily short coherence time. With this we indicate that, although it requires strict synchronization and proper power allocation, multiuser parametric channel estimation is a viable alternative for sensing and communication with \ac{OFDM} waveforms in intense Doppler environments. Extending the proposed algorithm for near-field and nonstationary channels is a promising direction for future work.
\section*{Acknowledgements}
The work was supported in part by the Research Council of Finland (former Academy of Finland) 6G Flagship Program (Grant Number: 346208) and 6GWiCE project (357719).

\ifCLASSOPTIONcaptionsoff
  \newpage
\fi



%\bibliographystyle{IEEEtran}

\bibliographystyle{IEEEtran}
\bibliography{IEEEabrv,biblio}

%\begin{thebibliography}{1}

%\end{thebibliography}


%\end{IEEEbiography}


%\newpage


%\vfill


\end{document}